\documentclass[aps,pra,twocolumn]{revtex4}
\usepackage{graphicx}
\usepackage{bm}
\usepackage{amsmath,amssymb,amsthm,dsfont,bm}
\usepackage{color}
\usepackage{wasysym}
\usepackage{ulem}
\usepackage{appendix}
\usepackage[dvipsnames]{xcolor}

\usepackage[english]{babel} 
\usepackage[utf8]{inputenc} 
\usepackage{hyperref}

\begin{document}
\preprint{}
\title{Coherence, nonclassicality and metrological resolution }
\author{Laura Ares}
\email{laurares@ucm.es}
\author{Alfredo Luis}
\email{alluis@fis.ucm.es}
\homepage{http://www.ucm.es/info/gioq}
\affiliation{Departamento de \'{O}ptica, Facultad de Ciencias
F\'{\i}sicas, Universidad Complutense, 28040 Madrid, Spain}
\date{\today}

\begin{abstract}
We demonstrate the simple and deep equivalence between quantum coherence and nonclassicality and the definite way in which they determine metrological resolution. Moreover, we define a coherence observable consistent with a classical intuition relating coherence with phase statistics.  
\end{abstract}
\maketitle

\section*{Introduction}

Coherence is the physical principle by which quantum mechanics differs from classical mechanics and classical optics is more than geometrical optics. Here we focus on the quantum side, so we deal with the fact that coherence is the source of nonclassicality and its applications, quantum metrological resolution in particular \cite{SP17,TJ19}. 

\bigskip
 
\noindent {\it Definition of coherence.--} Coherence is always a base-dependent property. For definiteness let us always consider a finite-dimensional space of dimension $N$ and denote as $|j\rangle $, for integer $j=1,\ldots, N$, an orthonormal basis that defines a given observable $G$. For any state represented by its density matrix $\rho$ the {\it coherences} relative to $G$ are all its nondiagonal matrix elements
\begin{equation}
\label{ct}
    \langle j | \rho | k \rangle , \quad j \neq k .
\end{equation}
On the other hand, the diagonal terms provide the statistics of $G$
\begin{equation}
\label{dt}
    p_j = \langle j | \rho | j \rangle .
\end{equation}

\bigskip

\section*{Coherence equals nonclassicality}

\noindent {\it Definition of nonclassicality.--} 
Nonclassicality is always revealed by the lack of a proper joint distribution for two or more incompatible observables. The paradigmatic example in quantum optics is provided by the lack of a {\it bona fide} Glauber-Sudarshan $P(\alpha)$ distribution, which attempts to be a joint distribution for the two field quadratures, meaning that $P(\alpha)$ cannot be a probability distribution \cite{MW95}. Quadrature squeezing and subPoissonian photon statistics  are nonclassical evidences because they are incompatible with a  {\it bona fide} $P(\alpha)$. But nonclassicality goes far this example, and we may recall the simple and ready to use approach presented in Refs. \cite{AL16a,AL16b,LM17}.

\bigskip

So, besides $G$ we must invoke another observable, say $M$, not compatible with $G$, this is no jointly measurable. With this in mind Let us demonstrate the quantum nature of $\rho$ with respect to $G$ lies on the non-diagonal matrix elements of $\rho$ in the basis $G$. 

\bigskip

\noindent {\it Theorem.--} $\rho$ is nonclassical if and only if there is a basis $G$ such that  $\langle j | \rho | k \rangle \neq 0$ for at least a pair of basis elements $j\neq k$.

\noindent {\it Proof.--} To prove that coherence implies nonclassicality we restrict ourselves to the two-dimensional subspace spanned by the basis components $|j\rangle$ and $|k \rangle$ involved in a non vanishing coherence term, where we can define the Pauli-like matrices
\begin{eqnarray}
    & \sigma_z = |j \rangle \langle j| - |k \rangle \langle k| , & \nonumber \\
& \sigma_\phi = e^{i \phi}|j \rangle \langle k| + e^{-i \phi}|k \rangle \langle j| , & \\
& \sigma_{\perp \phi} = i (e^{i \phi}|j \rangle \langle k| - e^{-i \phi}|k \rangle \langle j|) , & \nonumber
\end{eqnarray}
where $\phi$ is chosen so that
\begin{equation}
    \langle \sigma_\phi \rangle = 2 |\langle j | \rho | k \rangle|, \quad  \langle \sigma_{\perp \phi} \rangle = 0.
\end{equation}

To address nonclassicality we apply the approach in Refs. \cite{AL16a,AL16b,LM17} that begins with a joint noisy measurement of $\sigma_z$, say $G$, and $\sigma_{\perp \phi}$, say $M$, and performs a data inversion leading to a noise-free joint distribution for $\sigma_z$ and $\sigma_{\perp \phi}$, which results in the form
\begin{equation}
\label{pyz}
    p (y,z) \propto 1+ z \langle \sigma_z \rangle + y \langle \sigma_{\perp \phi} \rangle +  yz \frac{\gamma_{yz}}{\gamma_y \gamma_z} \langle \sigma_\phi \rangle, 
\end{equation}
where $y,z=\pm 1$ are dichotomic variables and the $\gamma$ are parameters characterizing the measurement, which are constrained to obey the relation 
\begin{equation}
    \gamma_y^2+\gamma_z^2+\gamma_{yz}^2 \leq 1 .
\end{equation}
We end the implication noting that whenever $|\langle j | \rho | k \rangle|\neq 0$, this is $\langle \sigma_\phi \rangle \neq 0$, we can always choose the $\gamma$ factors and the $y,z$ values so that  $p(y,z)<0$, for example by considering $\gamma_{yz}\neq 0$ and small enough $\gamma_y \gamma_z$. 

To prove the converse so that nonclassicality implies coherence we note that the only way that $p(y,z)\geq 0$ in Eq. (\ref{pyz}) for all $y,z$ and all the allowed $\gamma$ values is that 
\begin{equation}
      \langle \sigma_\phi \rangle =   0, \rightarrow \langle j | \rho | k \rangle = 0.  \qquad \blacksquare
\end{equation}

\bigskip

\noindent {\it Comment.--} Let us note that $\rho$ is nonclassical with respect to $G$ provided that $[G,\rho] \neq 0$, which agrees with the recent result in Ref. \cite{SL17}. So we may say that deep down the very same $\rho$ is the observable $M$ that reveals its nonclassicality with respect to $G$ when $[G,\rho] \neq 0$.

\bigskip

\noindent {\it Corollary.--} Every state $\rho \neq I/N$, where $I$ is the identity matrix, is nonclassical.  This is because the identity is the only matrix that is diagonal in all the bases. Otherwise, if for example $\langle j | \rho | j \rangle \neq  \langle k | \rho | k \rangle$ for a pair $j,k$ we can always perform a basis change in the subspace spanned by $|j\rangle$, $|k \rangle$, i. e., a rotation in the corresponding Bloch sphere, so that a nonvanishing coherence term appears.

\bigskip

\noindent {\it Comment.--} The above corollary is the very same reason by which for two field modes the degree of polarization is an upper bound for their degree of coherence \cite{BW99}.  

\bigskip

\section*{Coherence/nonclassicality as an observable }

Let us proceed developing further the idea that coherence/nonclassicality is a multiple-observable property by constructing the coherence observable $\Phi$ relative to some observable $G$. 

\bigskip

\noindent  {\it Definition of the coherence observable.-} To this end we define the following unnormalized and nonorthogonal phase states in the basis $G$  \cite{AL08,AL09,AL10}
\begin{equation}
\label{mps}
    |\bm{\phi} \rangle = \sum_{j=1}^N e^{i \phi_j} |j \rangle, 
\end{equation}
in terms of $N$ independent phase variables
\begin{equation}
    \bm{\phi} = (\phi_1,\phi_2,\ldots,\phi_N ),
\end{equation}
taking all them any value in a  $2\pi$ interval. Actually, since a global phase is irrelevant, we may reduce it to $N - 1$ phases, but we keep $N$ for the sake of symmetry and simplicity. These vectors allows us to construct a positive operator-valued measure (POVM) representing the multi-phase observable $\Phi$ as
\begin{equation}
    \Delta (\bm{\phi}) = | \bm{\phi} \rangle \langle \bm{\phi}|, 
\end{equation}
defining a multi-phase probability distribution 
\begin{equation}
P(\bm{\phi}) = \langle \bm{\phi} | \rho |\bm{\phi} \rangle = \sum_{j,k=1}^N  e^{i (\phi_k - \phi_j)}\langle j | \rho | k \rangle ,
\end{equation}
this is 
\begin{equation}
\label{mpp}
P(\bm{\phi}) = 1 + \sum_{ j \neq k=1}^N e^{i (\phi_k - \phi_j)}\langle j | \rho | k \rangle ,
\end{equation}
with a phase measure $d\phi$ 
\begin{equation}
d \bm{\phi} = \frac{\Pi_{j=1}^N d\phi_j }{(2\pi)^N} ,\quad \int_{2\pi} d\phi = 1 , 
\end{equation}
so that 
\begin{equation}
\int_{2\pi} d\phi P(\bm{\phi}) =1.
\end{equation}

\bigskip
\noindent {\it Properties of $\Phi$.-} Let us present some interesting properties of $\Phi$ that support the idea that it represents the coherence observable on its most general form.

\bigskip
i.- As it can be clearly seen in Eq. (\ref{mpp}) $P ( \bm{\phi})$ only depends on the coherence terms (\ref{ct}) and does not depend on the diagonal terms (\ref{dt}).

\bigskip
ii.- There is a clear complementarity between $\Phi$ and $G$ since  $| \langle j |\bm{\phi} \rangle | = 1$. 

\bigskip
iii.- $P ( \bm{\phi})$  contains the same information that all coherence terms (\ref{ct}),
\begin{equation}
    \int d\phi e^{i (\phi_j -\phi_k)} P ( \bm{\phi}) =  \langle j | \rho | k \rangle, \quad j\neq k .
\end{equation}

\bigskip
iv.- $\Delta ( \bm{\phi})$ is covariant under the phase shifts generated by any diagonal infinitesimal generator $|j \rangle \langle j |$, this is to say 
\begin{equation}
    P(\phi_j ; U \rho U^\dagger ) =  P(\phi_j + \lambda ; \rho )  , \quad U= e^{-i \lambda |j \rangle \langle j |} ,
\end{equation}
without any change in the other phase variables $\phi_{k\neq j}$. This will be useful  when addressing the connection between coherence/nonclassicality and quantum metrology. 

\bigskip

The property iii means that any measure of coherence introduced so far is expressible in terms of the statistics $P ( \bm{\phi})$ of the coherence observable $\Phi$. In particular, this must be the case of the Hilbert-Schmidt coherence
\begin{equation}
\label{HScm}
    \mathcal{C}_{HS} = \sum_{j\neq k=1}^N \left | \langle j | \rho | k \rangle \right |^2 .
\end{equation}

\bigskip

\noindent {\it Theorem.-} The Hilbert-Schmidt measure of coherence (\ref{HScm}) is given by the R\'enyi entropy of the coherence observable $\Phi$ \cite{AR70}:
\begin{equation}
\label{P2CHS}
 \int_{2\pi} d\phi P^2(\bm{\phi}) = 1 + \mathcal{C}_{HS} .
\end{equation}

\noindent {\it Proof.-} The demonstration follows form the equality 
\begin{equation}
\label{P2}
 P^2(\bm{\phi})= \sum_{j,k,\ell,m=1}^N e^{i (\phi_k - \phi_j + \phi_m - \phi_\ell)}\langle j | \rho | k \rangle \langle \ell | \rho | m \rangle ,
\end{equation}
where there are only two types of nonvanishing terms. The first one holds for 
\begin{equation}
   k = j, \quad  m = \ell ,
\end{equation}
leading to the contribution
\begin{equation}
\label{1c}
\sum_{j,\ell=1}^N \langle j | \rho | j \rangle \langle \ell | \rho | \ell \rangle =1 .
\end{equation}
The second type holds for 
\begin{equation}
  k =\ell, \quad  m = j ,
\end{equation}
excluding the case $k=j$ already included in the first case, leading to the contribution 
\begin{equation}
\label{2c}
\sum_{j \neq k=1}^N \langle j | \rho | k \rangle \langle k| \rho | j \rangle = \sum_{j \neq k=1}^N | \langle j | \rho | k \rangle |^2 = \mathcal{C}_{HS} .  \end{equation}
So, Eqs. (\ref{P2}), (\ref{1c}) and (\ref{2c}) lead to (\ref{P2CHS}).  $\blacksquare$

\bigskip

\section*{Coherence/nonclassicality and resolution}

The Hilbert-Schmidt coherence-measure is one of many abstract possibilities that puts the same weight on all coherence terms (\ref{ct}). But in real practical applications of coherence, such as quantum metrology, not all coherence terms will contribute equally. So that the relation between coherence and metrological resolution may be clear and definite, although not expressible directly in terms of standard coherence measures as (\ref{HScm}). Let us dwell on this point by means of some examples. Throughout we consider that some signal $\lambda$ is imprinted on the field state $\rho$ by means of the unitary transformation 
\begin{equation}
\label{g}
     U (\lambda)= e^{-i \lambda g } , \quad g = \sum_{j=1}^N j |j \rangle \langle j | .
\end{equation}
\bigskip

\noindent{\it Wiener-Kintchine theorem.--} After Eq. (\ref{g}) there is a clear definite relation between the phase shifts experienced by all phases $\phi_j$. So all the information about the signal can be more simply obtained considering a marginal version of the POVM $\Delta (\bm{\phi})$ obtained after the following constraints on the phase-coherence variables 
\begin{equation}
\label{pc}
\phi_j = j \phi , 
\end{equation} 
which is is equivalent to consider the observable POVM 
\begin{equation}
\label{Dp}
    \Delta (\phi) = \frac{1}{2 \pi} | \phi \rangle \langle \phi |, 
\end{equation}
where $| \phi \rangle$ is derived from (\ref{mps}) after the constraints (\ref{pc}), so that 
\begin{equation}
    |\phi \rangle = \sum_{j=1}^N e^{i j \phi} | j \rangle.
\end{equation}
Naturally, the so defined single-phase POVM  $\Delta(\phi)$ inherits the covariance of the original multi-phase POVM $\Delta (\bm{\phi})$ in the form
\begin{equation}
    U^\dagger (\lambda) \Delta(\phi) U (\lambda) = \Delta(\phi + \lambda).
\end{equation}
This is an example of quantum ruler as introduced in Ref. \cite{LA20} so we can apply the quantum Wiener-Kintchine theorem establishing a quite direct link between resolution and quantum coherence in the form  \cite{LA20}
\begin{equation}
\Delta^2 \lambda = \frac{1}{2 \sqrt{\pi} \int d\phi P^2 (\phi)}  ,
\end{equation}
where
\begin{equation}
\label{PG}
    P(\phi) = \frac{1}{2 \pi} \sum_{j,k=1}^N \langle j |\rho | k \rangle e^{i(k-j) \phi} = \sum_{\tau=-N}^N \Gamma (\tau ) e^{-i\tau \phi} ,
\end{equation}
where $\tau$ is integer and $\Gamma (\tau )$ is the mutual coherence function 
\begin{equation}
   \Gamma (\tau ) = \frac{1}{2 \pi} \sum_{j=1}^N
   \langle j+\tau |\rho | j \rangle,
\end{equation}
and it is understood that $|k\rangle$ vanishes if the index $k$ is out of the range  $k=1,2,\ldots , N$. This can be also expressed as 
\begin{equation}
   \Gamma (\tau )= \mathrm{tr} \left ( \rho E^\tau \right ) ,
\end{equation}
where $E$ is the analog Susskind-Glogower exponential-of-phase operator \cite{qop}, this is 
\begin{equation}
    E= \sum_{j=1}^N | j \rangle \langle j+1 | ,
\end{equation}
and $E^\tau$ for $\tau <0$ means $(E^\dagger)^{|\tau|}$. 
This agrees with some previous works formulating coherence as phase statistics \cite{AL09}. 

Using Eq. (\ref{PG}) we get that 
\begin{equation}
    \int d\phi P^2 (\phi) = 2 \pi \sum_{\tau=-N}^N | \Gamma (\tau ) |^2, 
\end{equation}
which from the perspective of the Wiener-Kintchine classical coherence theorem can be regarded as a measure of the coherence time as the area under the coherence function $| \Gamma (\tau ) |$ normalized so that $| \Gamma (\tau=0 ) | =1/(2 \pi)$. We have that lesser coherence time means lesser area and vice versa. Note also that at difference with other measures of coherence this depends on the phases of the coherences (\ref{ct}) as well as on their modulus.

\bigskip

Beyond this coherence-resolution theory, in general there is a lack of direct relation of resolution with the coherence measures introduced so far, including some of the recently introduced in terms of Fisher information rigorously respecting resource theories \cite{LL21}. This is because of the factors $g_j$ for the most general generator we have 
\begin{equation}
\label{mgg}
     U (\lambda)= e^{-i \lambda g } , \quad g = \sum_{j=1}^N g_j |j \rangle \langle j | ,
\end{equation}
and no theory based just on the coherence terms (\ref{ct}) takes into account the particular factors $g_j$, unless they are more or less trivial as in Eq. (\ref{g}).  

\bigskip
\bigskip

\noindent{\it Statistical distance.--} We can measure the resolution of a signal-detection scheme by the effect of the signal on the measured  statistics, for example in terms of a Hilbert-Schmidt distance 
\begin{equation}
\label{dHS0}
    d^2_{HS} = \sum_\mu \left [ p_\mu (\lambda ) - p_\mu (\lambda=0 ) \right ]^2 ,  
\end{equation}
where $p_\mu (\lambda )$ is the statistics of the measurement described by some POVM $\Delta (\mu )$, and $\mu$ are its outcomes. For small enough signal $\lambda$ this scales as 
\begin{equation}
\label{dHS}
    d^2_{HS} \simeq  \lambda^2 \sum_\mu p^{\prime 2}_\mu , \quad    p^{\prime}_\mu = \left . \frac{d}{d \lambda} p_\mu (\lambda ) \right |_{\lambda =0} , 
\end{equation}
which leads to 
\begin{equation}
  p^{\prime}_\mu = i \sum_{j\neq k=1}^N \left (g_k - g_j \right ) \langle j | \rho | k \rangle  \langle k | \Delta (\mu ) | j \rangle .
\end{equation}

\bigskip

We can appreciate that this is a valid measure of resolution directly expressible just in terms of the coherence terms (\ref{ct}) that does not depend on the diagonal matrix elements (\ref{dt}). Moreover, $d_{HS}$ clearly includes the contribution of the coherences of the detector, in a similar way we found in the quantum Wiener-Kintchine theorem in Ref. \cite{LA20}.

\bigskip

Furthermore, using standard variance-based uncertainty relations applied to the infinitesimal generator $g$ in Eq. (\ref{mgg}) leads to the following bound 
 \begin{equation}
  4 \Delta^2 g  \geq \frac{\sum_\mu p^{\prime 2}_\mu}{\eta - \sum_\mu p^2_\mu}  ,
 \end{equation}
where $\Delta^2 g$ is the variance of the infinitesimal generator $g$ on the probe state, while the factor $\eta$ comes for the following assumption regarding the measurement POVM, $[\Delta (\mu )]^2 = \eta \Delta (\mu )$. 

We can see that the right-hand side is a kind of signal/noise resolution measure. In the numerator we have essentially the statistical distance $d_{HS}$ between the signal-transformed and original probability distributions in Eq. (\ref{dHS}). In the denominator there is essentially the uncertainty of the original probability distribution measured in terms of a Renyi entropy \cite{AR70}. Finally, the upper bound in the left-hand side is also clearly recognizable as the quantum Fisher information.
 
\bigskip

\noindent{\it Density-matrix distance.--} We may consider as well an intrinsic resolution measure similar to Eq. (\ref{dHS0}) that compares density matrices instead of probabilities \cite{RL08}
\begin{equation}
    D^2_{HS} = \mathrm{tr} \left [ \rho (\lambda ) - \rho (\lambda=0 ) \right ]^2 ,  
\end{equation}
where 
\begin{equation}
    \rho (\lambda ) = U(\lambda) \rho (\lambda=0 ) U^\dagger (\lambda) ,
\end{equation}
and $U(\lambda )$ is in Eq. (\ref{mgg}). Again, for small enough signal this scales as 
\begin{equation}
\label{DHS}
    D^2_{HS} \simeq -  \lambda^2 \mathrm{tr} \left ( [ \rho ,g ]^2 \right ) ,
\end{equation}
where simply $\rho = \rho (\lambda=0 )$ which leads to 
\begin{equation}
    D^2_{HS} \simeq  \lambda^2 \sum_{j\neq k=1}^N \left (g_k - g_j \right )^2 | \langle j | \rho | k \rangle |^2 
\end{equation}
depending just on the coherence terms (\ref{ct}) being independent of the  diagonal matrix elements (\ref{dt}). 

\bigskip

\section*{Conclusions}
We have elaborated a simple and complete equivalence between nonclassicality and coherence in terms of basic and universal concepts, beyond particular measures and assessments of these magnitudes. We have shown that they are one and the same concept. 

Moreover this fully determines a basic application of coherence as it is metrological resolution. Deep down, this was actually the origin of coherence as introduced in classical optics to properly account for interferometric visibility. 

In this regard, we have also translated to the quantum domain the classic intuition regarding partial coherence in classical optics as the result of phase fluctuations. This also confirms a basic intuition regarding coherence as the actual source of quantum phenomena.   
\bigskip

\section*{ACKNOWLEDGMENTS}
L. A. and A. L. acknowledge financial support from Spanish Ministerio de Econom\'ia y Competitividad Project No. FIS2016-75199-P.
L. A. acknowledges financial support from European Social Fund and the Spanish Ministerio de Ciencia Innovaci\'{o}n y Universidades, Contract Grant No. BES-2017-081942.

\end{document}